\documentclass{elsart}
\usepackage[dvips]{graphicx}

\begin{document}
\begin{frontmatter}

\title{ Limits on different Majoron decay modes of $^{100}$Mo and $^{82}$Se
 for neutrinoless double beta decays in the NEMO-3 experiment}

\author[IReS]{R.~Arnold},
\author[LAL]{C.~Augier},
\author[INEL]{J.~Baker},
\author[ITEP]{A.S.~Barabash$^{\;1)}$}\thanks{Corresponding author, 
Institute of Theoretical and Experimental Physics, B.~Cheremushkinskaya 25,
117259 Moscow, Russia,  e-mail: Alexander.Barabash@itep.ru,
tel.: 007 (095) 129-94-68, fax: 007 (095) 883-96-01},
\author[JINR]{V.~Brudanin},
\author[INEL]{A.J.~Caffrey},
\author[IReS]{E.~Caurier},
\author[JINR]{V.~Egorov},
\author[LAL]{K.~Errahmane},
\author[LAL]{A.I.~Etienvre},
\author[IReS]{J.L.~Guyonnet},
\author[CENBG]{F.~Hubert},
\author[CENBG]{Ph.~Hubert},
\author[IReS]{C.~Jollet},
\author[LAL]{S.~Jullian},
\author[UCL]{S.~King},
\author[JINR]{O.~Kochetov},
\author[ITEP]{S.~Konovalov},
\author[JINR]{V.~Kovalenko},
\author[LAL]{D.~Lalanne},
\author[CENBG]{F.~Leccia},
\author[LPC]{C.~Longuemare},
\author[CENBG]{G.~Lutter},
\author[IReS]{Ch.~Marquet},
\author[LPC]{F.~Mauger},
\author[IReS]{F.~Nowacki},
\author[SAGA]{H.~Ohsumi}
\author[CENBG]{F.~Piquemal},
\author[CFR]{J-L.~Reyss},
\author[UCL]{R.~Saakyan},
\author[LAL]{X.~Sarazin},
\author[JINR]{Yu.~Shitov},
\author[LAL]{L.~Simard},
\author[FMFI]{F.~\v{S}imkovic},
\author[JINR]{A.~Smolnikov},
\author[CTU]{I.~\v{S}tekl},
\author[JYV]{J.~Suhonen},
\author[MHC]{C.S.~Sutton},
\author[LAL]{G.~Szklarz},
\author[JINR]{V.~Timkin},
\author[UCL]{J.~Thomas},
\author[JINR]{V.~Tretyak},
\author[ITEP]{V.~Umatov},
\author[CTU]{L.~V\'{a}la},
\author[ITEP]{I.~Vanyushin},
\author[ITEP]{V.~Vasilyev},
\author[CU]{V.~Vorobel},
\author[JINR]{Ts.~Vylov}
\address[CENBG]{CENBG, IN2P3-CNRS et Universit\'e de Bordeaux,
               33170 Gradignan, France}
\address[LPC]{LPC, IN2P3-CNRS et Universit\'e de Caen, 14032 Caen, France}
\address[JINR]{JINR, 141980 Dubna, Russia }
\address[CFR]{CFR, CNRS, 91190 Gif sur Yvette, France}
\address[INEL]{INL, Idaho Falls, ID 83415, U.S.A.}
\address[JYV]{JYV\"ASKYL\"A University, 40351 Jyv\"askyl\"a, Finland}
\address[ITEP]{ITEP, 117259  Moscow, Russia}
\address[LAL]{LAL, IN2P3-CNRS et Universit\'e Paris-Sud, 91405 Orsay, France}
\address[MHC]{MHC, South Hadley, Massachusetts 01075, U.S.A.}
\address[IReS]{IReS, IN2P3-CNRS et Universit\'e Louis Pasteur, 67037 Strasbourg, France.}
\address[CTU]{IEAP, Czech Technical University, CZ-128 00 Prague, Czech Republic.}
\address[CU]{Charles University, Prague, Czech Republic.}
\address[SAGA]{SAGA University, Saga, Saga 840-8502, Japan}
\address[UCL]{University College London, London, UK }
\address[FMFI]{FMFI, Comenius University,SK-842 48 Bratislava, Slovakia}
{\large NEMO Collaboration}
\date{ }

\begin{abstract}
The NEMO-3 tracking detector is located in the Fr\'ejus Underground
Laboratory.  It was designed to study double beta decay in a number of 
different isotopes.
Presented here are the experimental half-life limits on the double beta decay 
process for the isotopes
$^{100}\rm Mo$ and $^{82}\rm Se$
for different Majoron emission modes and limits on the effective
neutrino-Majoron coupling constants. In particular, 
new limits on "ordinary" Majoron 
(spectral index 1) decay of $^{100}\rm Mo$ ($T_{1/2} > 2.7\cdot10^{22}$ y) 
and $^{82}\rm Se$ ($T_{1/2} > 1.5\cdot10^{22}$ y)
have been obtained. Corresponding bounds on the Majoron-neutrino 
coupling constant 
are $\langle g_{ee} \rangle < (0.4-1.9) \cdot 10^{-4}$ 
and  $< (0.66-1.7) \cdot 10^{-4}$.

{\it PACS:} 23.40.-s, 14.80.Mz

\begin{keyword}
Majoron, double-beta decay.
\end{keyword}
\end{abstract}
\end{frontmatter}

\renewcommand{\baselinestretch}{1.6}
\newpage

\section{Introduction}
Spontaneous violation of global (B-L) symmetry in gauge theories
leads to the existence of a massless Goldstone boson, the Majoron. The Majoron,
if it exists, 
could play an important role in Cosmology \cite{BER93,DOL04,KAZ04} and 
Astrophysics \cite{KAC00,TOM01,HAN02,FAR03}. At the
beginning of the 1980's, the
singlet \cite{Mohapatra80}, doublet \cite{Mohapatra82} and
triplet \cite{Gelmini81} Majoron models were developed. 
All these models resulted in the 
neutrinoless double beta  
decay with the emission of a Majoron 

\begin{equation}
(A,Z) \rightarrow (A,Z+2) + 2e^{-} + \chi^{0}
\end{equation}

However, the interaction of the triplet (or doublet) 
Majorons with the $Z^{0}$ boson
would give a contribution to the width of the $Z^{0}$ decay, 
which corresponds to two (or 1/2)
additional massless neutrino types
(see for example \cite{Georgi81,Barger82,Deshpande}).
LEP data gives $2.994 \pm 0.012$ neutrino
types \cite{Caso98}, thus the triplet and some doublet Majorons are excluded.
On the other hand the singlet "see-saw" Majoron \cite{Mohapatra80}
is extremely weakly coupled with neutrinos. 
Nevertheless, in reference \cite{Berezhiani92} it is proposed that a
small gauge coupling
constant  (which determines the Majoron coupling to the $Z^0$ boson) does not
eliminate the possibility of a 
large Yukawa coupling of Majoron to neutrinos.
Thus, the singlet or dominantly singlet Majorons can still
contribute to neutrinoless 2$\beta$ decay \cite{Berezhiani92,Mohapatra91}.

Another possibility  for neutrinoless 2$\beta$-decay with
Majoron emission arises in supersymmetric models
with R-parity violation \cite{Mohapatra91,Mohapatra88}.
Mohapatra and Takasugi~\cite{Mohapatra88} proposed that
there could be $2\beta\chi^{0}\chi^{0}$-decay
with the emission of two Majorons :

\begin{equation}
(A,Z) \rightarrow (A,Z+2) + 2e^{-} + 2\chi^{0}
\end{equation}

In the 1990's several new "Majoron" models were suggested.
The term "Majoron" here denotes
massless or light bosons with a coupling to neutrinos.
In these models the "Majoron" can carry a lepton charge, but cannot be
a Goldstone boson \cite{Burgess94}.
Additionally there can be decays with the emission of two "Majorons"
\cite{Bamert95}. In the models with a vector "Majoron", the Majoron is the
longitudinal component of the massive gauge boson emitted in $2\beta$ decay
\cite{Carone93}.
All these new objects are called Majorons for simplicity. The 
possibility for 2$\beta$ decay
with the emission of one or two Majorons was discussed also in the 
framework of the $SU(3)_L \otimes SU(1)_N$ 
electroweak model \cite{MON01}. 

Recently a new "economical" model for neutrino mass was proposed 
in the context of the brane-bulk scenarios for particle physics. 
In this model the standard global B-L symmetry is broken 
spontaneously by a gauge singlet Higgs field in the bulk. 
This leads to a bulk singlet Majoron whose Kaluza-Klein 
excitations may make it visible in neutrinoless double 
beta decay \cite{Moh00}.  

\begin{figure*}
\begin{center}
\resizebox{0.75\textwidth}{!}{\includegraphics{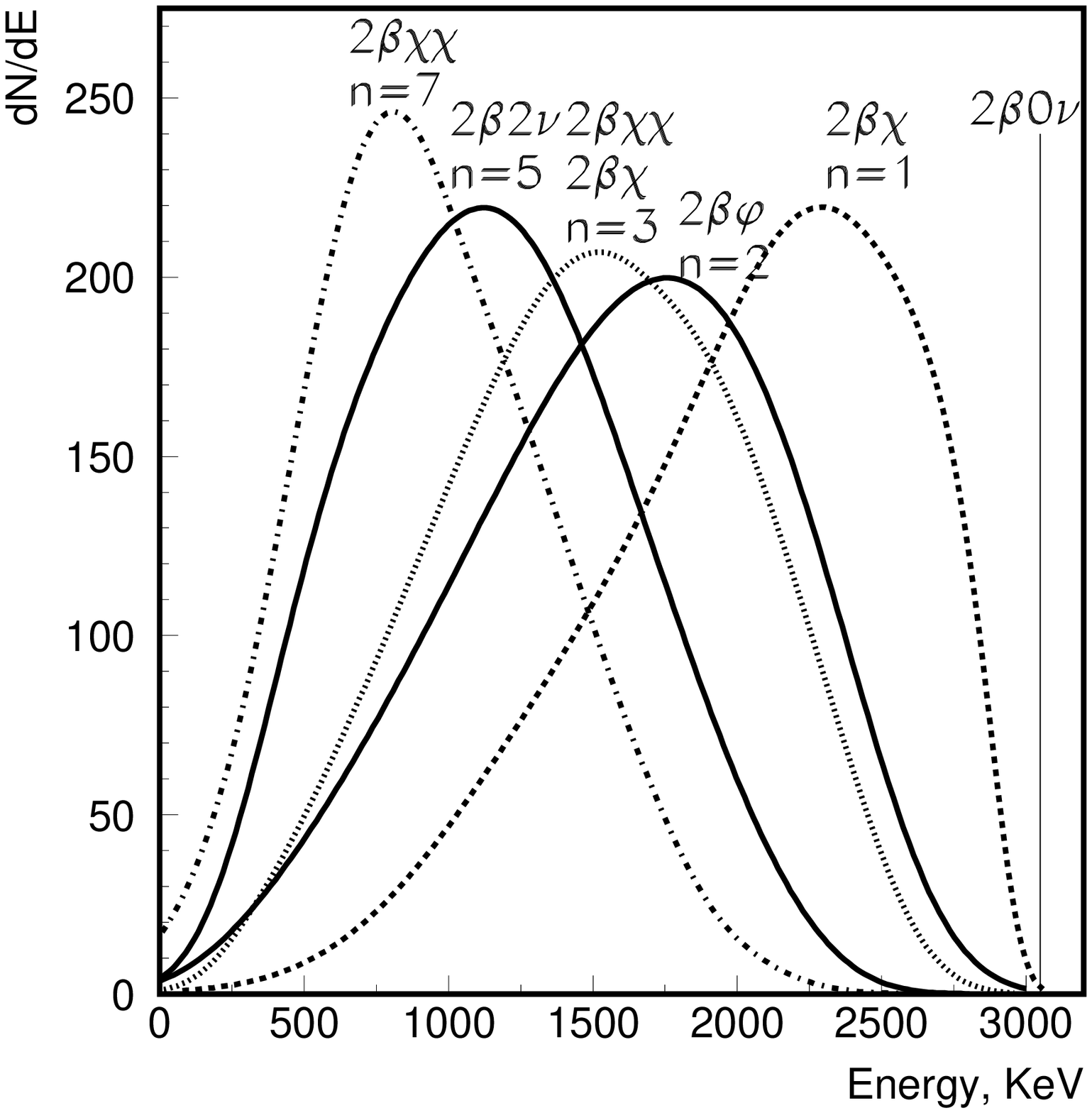}}
\caption{Energy spectra of different modes of $2\beta2\nu$ $(n=5)$,
$2\beta\chi^{0}$ $(n=1~$,$~2$ and$~3)$ and
$2\beta\chi^{0}\chi^{0} (n=3~$and$~7)$ decays of $~^{100}$Mo.}
\label{fig_modes}
\end{center}
\end{figure*}

In Table~1
there are 10 Majoron models presented (following 
\cite{Bamert95,Carone93,Moh00,Hirsch96}),
which are considered in this paper. 
It is divided into two sections, one for lepton
number violating (I) and one for lepton number 
conserving models (II). The last line corresponds to the bulk majoron model. 
The table also shows
whether the corresponding 2$\beta$ decay is accompanied by the emission of
one or two Majorons. The next three entries list the main
 features of the models:
the third column lists whether the Majoron is a Goldstone boson or not
(or a gauge boson in the case of vector Majorons, type IIF, 
or a bulk field Majoron).
In column four the leptonic charge L is given.
Column five gives the "spectral index" $n$ of the summed energy of the
emitted electrons, which is defined by
the phase space of the emitted particles, $G \sim (Q_{\beta\beta}-T)^{n}$.
Here $Q_{\beta\beta}$ is the energy released in the decay and $T$ the energy
of the two electrons. The energy spectra of different modes of
$2\beta2\nu$ $(n=5)$, $2\beta\chi^{0}$ $(n=1, 2~$and$~3)$
and $2\beta\chi^{0}\chi^{0}$ $(n=3~$and$~7)$ decays are presented in
Fig ~\ref{fig_modes}.
The different
shapes can be used to distinguish the different Majoron decay modes from
each other and 2$\beta$-decay with the emission of two neutrinos.
In the last column of Table~1
the nuclear matrix elements (NME) are listed.

Attempts to observe $2\beta$ decay with Majoron emission have been
carried out for the past 20 years. Consequently there now exist strong
limits on the "ordinary" Majoron with the "standard" electron energy 
spectrum shape
$(n=1)$, see Table~2.
Sufficiently less information exists for "non-ordinary" Majoron models.
The most carefully studied "non-ordinary" models are being 
investigated in \cite{Gunther96} for $~^{76}$Ge,  and in \cite{Arn678} for 
$^{100}\rm Mo$, $^{116}\rm Cd$, $^{82}\rm Se$ and $^{96}\rm Zr$ (see also \cite{Dan03} 
for $^{116}\rm Cd$). 

In this paper a systematic search for $2\beta$-decays with different 
Majoron types
is described for $^{100}$Mo and $^{82}$Se,
using the experimental data obtained with the NEMO-3 detector. 
The first results from NEMO-3 were published in  \cite{ARN04,BAR05,SAR05}.


\section{NEMO-3 detector}
\begin{figure*}
\begin{center}
\resizebox{0.75\textwidth}{!}{\includegraphics{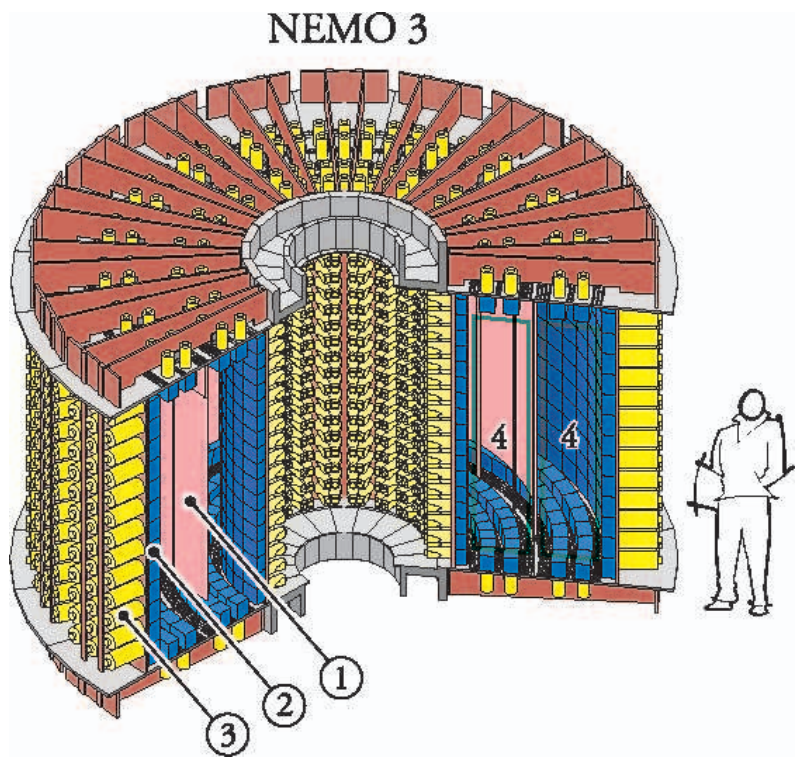}}
\caption{ The NEMO-3 detector without shielding. 1 -- source foil; 
2-- plastic scintillator; 3 -- low radioactivity PMT; 4 -- tracking chamber.}
\label{fig_NEMO3}
\end{center}
\end{figure*}

A schematic of the NEMO-3 detector is shown in 
(Fig.~\ref{fig_NEMO3}).
The main goal of the NEMO-3 experiment is to study neutrinoless 
double beta decay of different isotopes ($^{100}$Mo, $^{82}$Se etc.) 
with a sensitivity of up to $\sim 10^{25}$ y, which corresponds to 
a sensitivity to the effective Majorana neutrino mass at the level 
of $\sim(0.1-0.3)$ eV \cite{Proposal94}. The planned sensitivity to 
double beta decay with Majoron emission is $\sim 10^{23}$ y (the sensitivity 
to the coupling constant of the Majoron to the 
neutrino $<g_{ee}>$ is $\sim n\cdot10^{-5}$). 
In addition, one of the goals is a precise study of $2\beta2\nu$ decay for a 
number of nuclei ($^{100}$Mo, $^{82}$Se, $^{116}$Cd, $^{150}$Nd, $^{130}$Te, 
$^{96}$Zr and $^{48}$Ca) with high statistics to study all characteristics
 of the decay.

NEMO-3 is a tracking detector, which in contrast to $^{76}$Ge experiments 
\cite{Kla01,Aal02}, detects not only the total energy deposition but
also the path and energy of the individual electrons.
This also provides a unique opportunity to monitor and 
reject the background. Since June 2002 NEMO-3 has been running in the Fr\'ejus  
Underground Laboratory (France) located at a depth of 4800 m w.e. 

The detector has a cylindrical shape and consists of 20 identical 
sectors (see Fig~\ref{fig_NEMO3}). A thin ($\sim$30-60 mg/cm$^2$) source
foil placed in the center of the detector contains 2$\beta$ decaying nuclei 
and has a total area of 20 m$^2$ and a weight of about 10 kg. In particular, 
it includes 7.1 kg of enriched Mo (average enrichment 98\%, the total mass of 
$^{100}$Mo is 6.914 kg) and 0.96 kg of Se (enrichment 97\%, the total mass of 
$^{82}$Se is 0.932 kg). To investigate the external background the
part of the source 
are made of very pure natural material (TeO$_2$ -767 g and Cu - 621 g). 
The contamination of the sources with radioactive impurities was obtained from 
measurements using low-background HPGe-detectors.

The basic detection principles are the following: the energy of electrons 
is measured by plastic scintillators coupled to PMTs (1940 individual 
counters), while the tracks are reconstructed from the information obtained 
from the Geiger cells (6180 cells). The tracking volume of the detector 
is filled 
with a mixture consisting of 95\% He, 4\% alcohol, 1\% Ar and 0.15\% water 
at slightly above atmospheric pressure. In addition, a magnetic field of 25 Gauss parallel to the 
detector axis is created by a solenoid surrounding the detector. The magnetic 
field is used to identify electron-positron pairs and to suppress the 
background associated with these events.

The main characteristics of the detector's performance are the following: 
the energy resolution of the scintillation counters lies in the interval 
of 14-17\% (FWHM for 1 MeV electrons); the time resolution is 250 ps for an 
electron energy of 1 MeV; the reconstruction accuracy of a 2e$^-$  vertex 
is approximately 1 cm. The characteristics of the detector are determined in 
special calibration measurements with radioactive sources. The energy 
calibration is carried out using $^{207}$Bi sources (conversion electrons 
with energies 0.482 and 0.976 MeV) and $^{90}$Sr (the end-point of the 
$\beta$ spectrum is 2.283 MeV). The vertex reconstruction accuracy for 2e$^-$ 
events are determined by measurements with $^{207}$Bi, while the timing 
properties were determined in measurements with $^{60}$Co (two $\gamma$s 
emitted simultaneously), $^{207}$Bi (two electrons emitted 
simultaneously) and neutron sources (providing high energy electrons crossing 
the detector volume).

The detector is surrounded by passive shielding made of 20 cm of steel, 
30 cm of water in tanks around the detector and wood and paraffin at the top 
and bottom of the detector. The level of radioactive impurities in the 
construction materials of the detector and of the passive shielding was measured 
with low-background HPGe detectors.

A full description of the detector and its characteristics can be found in 
\cite{Arn536}. 

\section{Experimental data}
\subsection{$^{100}\rm Mo$}
\begin{figure*}
\begin{center}
\resizebox{0.75\textwidth}{!}{\includegraphics{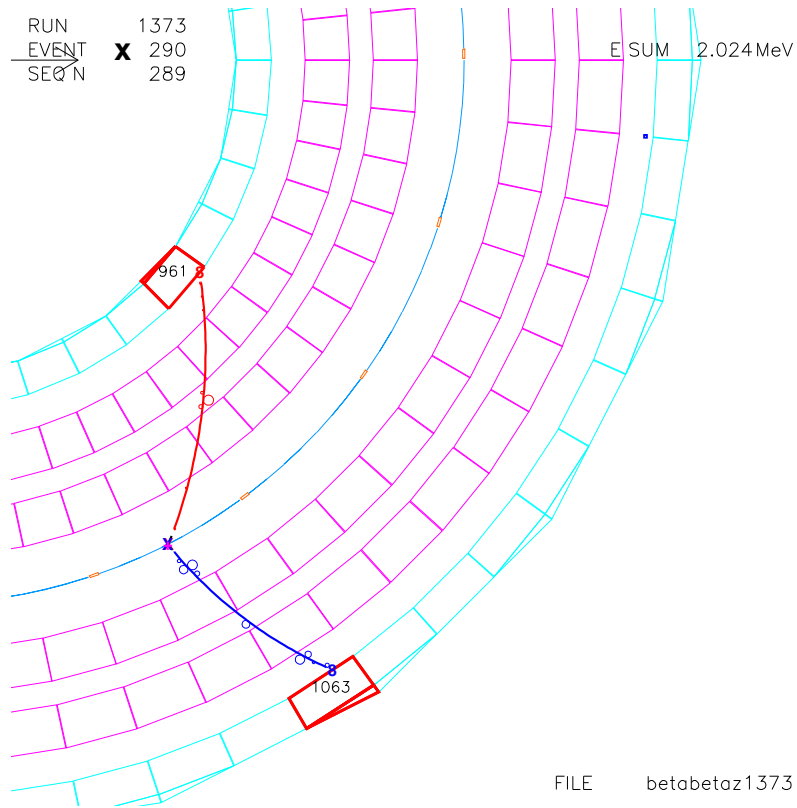}}
\caption{ A view of a reconstructed 2e event in NEMO-3. The sum energy 
of the electrons is 2024 keV; the energies of the electrons in the pair are 
961 keV and 1063 keV.}  
\label{fig_2eevent}
\end{center}
\end{figure*}

In this paper, the analysis of 8023 hours of NEMO 3 data is presented. 
2e events with a common vertex inside the source were selected. 
An electron  was defined as a track between the source 
foil and a fired counter with the energy deposited being greater than 
200 keV. The track curvature had to be consistent with a negatively charged 
particle; the time-of-flight measurement had to be consistent with the 
hypothesis of the two electrons leaving the source from a common vertex 
simultaneously. In order to suppress the $^{214}$Bi background, which is 
followed by the $^{214}$Po $\alpha$-decay, it was required that there were 
no delayed Geiger cell hits (with a delay of up to 700 $\mu$s) close to the 
event vertex or the electron track.  A typical 2e-event is shown in 
Fig~\ref{fig_2eevent}. 

\begin{figure*}
\begin{center}
\resizebox{0.75\textwidth}{!}{\includegraphics{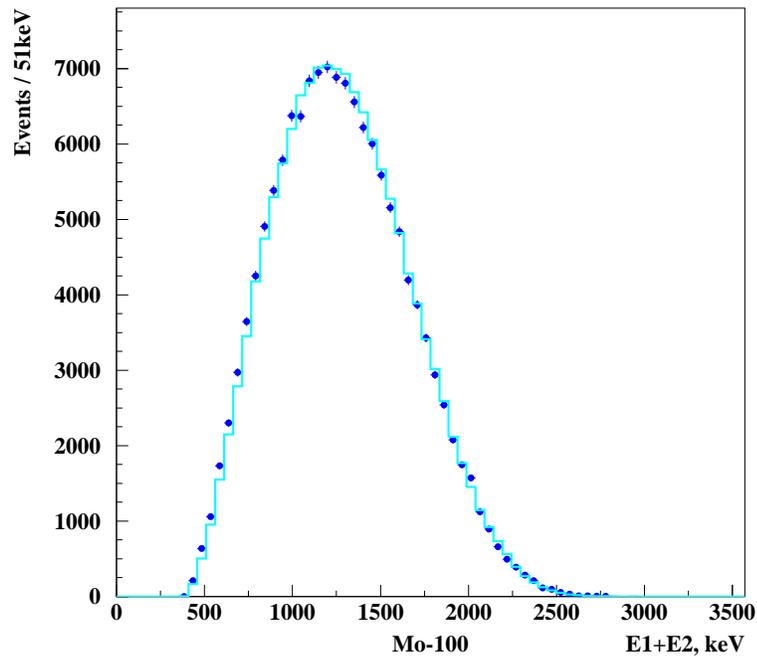}}
\resizebox{0.75\textwidth}{!}{\includegraphics{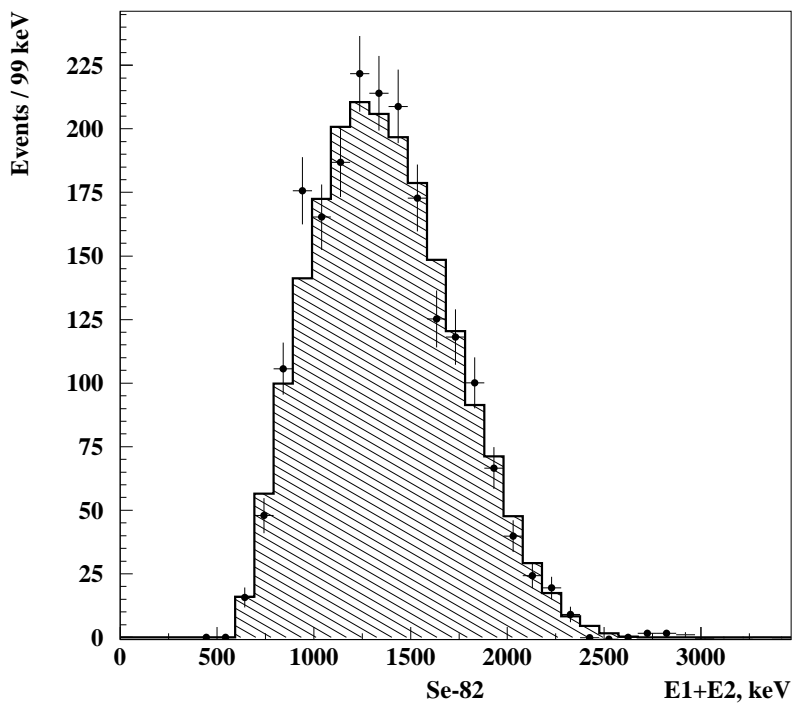}}
\caption{The 2e events (points - experiment; solid lines - Monte 
Carlo simulations for $2\beta2\nu$ decay) for  $~^{100}$Mo and
 $~^{82}$Se.}
\label{fig_data}
\end{center}
\end{figure*}

Fig~\ref{fig_data} (top) shows the $2\beta2\nu$ experimental energy spectrum 
and result of Monte Carlo (MC) simulations for $^{100}$Mo. 
The total number of useful events (after background subtraction) is 
$\sim$158000. The signal-to-background ratio is 40:1, while, for energies
above 1 MeV it is 100:1. 
This means that the background is negligibly small. 
The detection efficiencies which included the selection cuts were estimated 
for the single state dominance mechanism \cite{SIM01,DOM05} by 
MC simulations.  
The detection efficiency calculated by MC was 4.41\%. Correspondingly, 
the following results were obtained for the $^{100}$Mo half-life:

$$
  T_{1/2} = [7.41 \pm 0.02(stat) \pm 0.43(syst) ]\cdot 10^{18}\; y\  
$$

\subsection{$^{82}\rm Se$}
 The same 8023 h of data were analyzed. The experimental energy spectrum 
and result of MC simulations of  
$2\beta2\nu$ events for $^{82}$Se are shown in 
Fig~\ref{fig_data} (bottom). The total number of useful events after the background 
subtraction was $\sim2020$. The signal-to-background ratio was about 4:1. The 
detection efficiency was calculated by Monte Carlo to be 4.46\%. The $^{82}Se$ 
half-life value obtained is:

$$
 T_{1/2} = [9.6 \pm 0.24(stat)^{+0.67}_{-0.59} (syst) ] \cdot 10^{19}  y.		               
$$

This value is in agreement with the previous measurement made with the NEMO-2 
detector \cite{Arnold98}.

\section{Analysis of the Experimental Data}
The experimental data for $^{100}$Mo and $^{82}$Se are shown in
Fig.~\ref{fig_data}. One can see that 
experimental data are in a good agreement with the MC simulations. Exception is 
a low energy part of the $^{100}$Mo spectrum  (0.4-0.8 MeV), where the experimental points are 
systematically higher than the MC simulations. It can be associated with some physical effect 
(Majoron decay with $n=7$ or second-forbidden corrections contribution \cite{BAR99}, for example) or with 
our not ideal knowledge of the response function of the detector. To be conservative now
we prefer to be in framework of the last assumption.
 
The detection efficiencies for the decays
depend on the energy of the electrons and
were calculated for the two nuclei, for all the Majoron modes
(spectral indices $n=1,2,3~$and$~7$)
and for the double beta-decay $(n=5)$ by a
Monte-Carlo simulation with the GEANT 3.21 code.

If the Majoron modes are considered 
as existing decay channels similar to $2\beta 2\nu$, then the data
contains the sum of two processes, $2\beta 2\nu$ decay
and the decay with $\chi^{0}$ emission.
Therefore it is not possible to know the expected number of $2\beta 2\nu$ 
decays and so a limit must be set on the decays with Majoron emission
by analyzing the deviation in the shape of the energy distribution of
the experimental data in comparison with calculated spectrum
for $2\beta 2\nu$ decay. This can be done with a maximum likelihood analysis.

The experimental spectrum was treated as a histogram. One then needs to
take into account that the distribution of the events in each
bin is a Poisson one and independent of the others. Thus, one
constructs the likelihood function as:

\begin{equation}
\label{likelihood_formula}
\mathrm{L(N_{\beta},N_{\chi})} =
\prod_{\mathrm{i}=\mathrm{n_{1}}}^{\mathrm{n_{2}}}
\frac
{e^{-(\mathrm{ N_{\beta} \eta_{\beta~i} + N_{\chi} \eta_{\chi~i} +
N_{bgr~i}})}}
{\mathrm{N_{exp~i}}!}
(\mathrm{ N_{\beta} \eta_{\beta~i} + N_{\chi} \eta_{\chi~i} +
N_{bgr~i}})^{\mathrm{N_{exp~i}}} \,\,
\end{equation}

where $\mathrm{n_{1}}$ and $\mathrm{n_{2}}$ are the bin numbers of
the energy interval, $\mathrm{N_{exp~i}}$ is the number of experimental events
in the i-th bin, $\mathrm{N_{bgr~i}}$ is the expected number of background 
events, and
$\eta_{\beta~i}$ and $\eta_{\chi~i}$ are the
Monte-Carlo simulated efficiencies of $2\beta 2\nu$ and
Majoron decays in the i-th bin. Finally, $\mathrm{N}_{\beta}$ and
$\mathrm{N}_{\chi^{0}}$ are the average numbers of decays for $2\beta2\nu$ and
Majoran decay respectively, 
and are considered as free parameters.

To find the confidence level for the upper limit on the mean
number of decays with Majoron emission ($\mathrm{N}_{\chi up}$)
the likelihood function~(\ref{likelihood_formula}) has to be normalized
and then integrated over all possible values
of $\mathrm{N}_{\beta}$ and $\mathrm{N}_{\chi}$ from 0 to 
$\mathrm{N}_{\chi up}$:

\begin{equation}
\label{CL_likelihood}
\mathrm{CL(N_{\chi up})} =
\frac{
\int\limits_0^{\mathrm{N}_{\chi up}} \mathrm{dN}_{\chi}
\int\limits_0^{\infty} \mathrm{dN}_{\beta }
\mathrm{\: L(N_{\beta},N_{\chi})}
}
{
\int\limits_0^{\infty} \mathrm{dN}_{\chi}
\int\limits_0^{\infty} \mathrm{dN}_{\beta }
\mathrm{\:L(N_{\beta},N_{\chi})}
} \,\,
\end{equation}

where $\mathrm{N}_{\chi up}$ is a free parameter while
$\mathrm{CL}(\mathrm{N}_{\chi up})$ is fixed.

\section{Results and Discussion.}
The half-life limits for $^{100}$Mo and $^{82}$Se for the different decay modes
are presented in Table ~3.
For $~^{100}$Mo the limit on decays with $n=1$ and $n=2$  obtained here is 
$\sim$ 5 and $\sim$ 50 times higher than that in \cite{FUS02} and \cite{BAR04a}, 
with $n=3$
approximately 2 orders of magnitude
higher than that reported in \cite{Arn678}, and the limit on decays
with $n=7$ is improved only by a factor of  $\sim$ 1.5. 
In fact, in the latter case there are 
extra events in the low energy part of the spectrum and a conservative approach 
leads to a weak
limit on the half-life of this decay.
The result for $n=2$  is approximately 5 times better than 
estimated in \cite{BAR04a} while
the $~^{82}$Se results for n=1, 2 and 3 are improved over the ealier
limits \cite{Arn678,BAR04a} by $\sim 5-6$ times, and the limit for 
$n=7$ is improved by $\sim 50$ times.
Using the half-lives one can get limits on the coupling constants
for different Majoron models via the
relations (\ref{coupling2}) and (\ref{coupling4}).

\begin{equation}
\label{coupling2}
T_{1/2 }^{-1} = |\langle \mathrm{g}_{ee} \rangle|^{2}
|\mathrm{M}|^2 \mathrm{G} \,\, \mathrm{for \;\; 2\beta\chi^{0}} ,
\end{equation}

\begin{equation}
\label{coupling4}
T_{1/2}^{-1} = |\langle \mathrm{g}_{ee} \rangle|^{4}
|\mathrm{M}|^2 \mathrm{G} \,\, \mathrm{for \;\; 2\beta\chi^{0}\chi^{0}} ,
\end{equation}

The relevant matrix elements $\mathrm{M}$
and values of the phase space factors $\mathrm{G}$ are presented in 
Tables~4 and 5.
Using the data from Table~3
the limits on the coupling constants are calculated and
presented in Table~6.

The summary of the best  limits on the coupling constant of the Majoron to 
neutrinos for 
"ordinary" Majorons with n = 1 are presented in Table 2. 
One of the problems 
is the uncertainty in the Nuclear Matrix Element (NME) calculations 
which lead to a
dispersion of the $\langle g_{ee} \rangle$ value. 
In the $3^{rd}$ column, limits obtained using QRPA 
(different models) NME from \cite{SIM99,STO01,CIV03} are presented. Exceptions 
are $^{48}\rm Ca$ where Shell Model calculations have been used 
\cite{Retamosa95,Caurier99}
and $^{150}\rm Nd$ for which  NME values were taken from 
\cite{Hirsch95} where a Pseudo-SU(3) model taking into account the deformation
of the $^{150}\rm Nd$ nuclei was applied and from \cite{SIM99} were
calculations in the framework of QRPA were done 
(though such an approach is not really 
correct for deformed nuclei).

In the $4^{th}$ column of Table~2, limits using the NME from \cite{ROD05} 
are shown where the RQRPA model was used. 
In this recent work the suppression effect of higher order terms of 
the nucleon current have been taken 
into account and the $g_{pp}$ values were extracted 
from $2\beta2\nu$ experiments. The authors analyzed practically all 
the previous QRPA
and RQRPA calculations and concluded that their last calculations give the 
most reliable and accurate values for NME \footnote{This is not related 
to the $^{150}$Nd result, which is 
presented in \cite{ROD05} just for illustration owing to its deformed nuclear
shape}. 
If this is indeed the correct approach to the determination of  $g_{pp}$
then the best present limit is 
obtained from our measurements with $^{82}\rm Se$ and $^{100}\rm Mo$: 
$\langle g_{ee} \rangle < (1.2-1.9) \cdot 10^{-4}$ and 
$\langle g_{ee} \rangle < (1.6-1.8) \cdot 10^{-4}$ respectively.
If, however, the former approach is taken, and the 
NME values from 
\cite{SIM99,STO01,CIV03} are used, the best limit is obtained from the 
measurement of  $^{100}\rm Mo$ : $\langle g_{ee} \rangle < (0.4-0.7) \cdot 10^{-4}$. One can see from 
the Table 2 that new the approach leads to more conservative limits on the $\langle g_{ee} \rangle$ coupling 
constant for all nuclei.

All limits in Table 2  were obtained using phase space factors calculated in 
\cite{SUH98}. These values are 
$\sim$ 20\% lower than values 
obtained from \cite{SIM99} and the limits are therefore conservative and could
be a further 10\% more sensitive. 

To summarize briefly all the experimental results and taking 
into account uncertainties in NME calculations,
the conservative limit on $\langle g_{ee} \rangle$ from double beta decay 
experiments ("ordinary" Majoron) is at the level $ <2 \cdot 10^{-4}$. 
It is interesting to note that 
the Majoron-neutrino coupling constant in the 
range $ 4 \cdot 10^{-7}  < \langle g_{ee} \rangle  < 0.2 \cdot 10^{-4}$ is excluded 
by the observation of SN 1987A 
\cite{KAC00,FAR03}. This means that the possible range 
$ 2 \cdot 10^{-5}  < \langle g_{ee} \rangle  < 2 \cdot 10^{-4}$ is 
still allowed in contrast to conclusions from \cite{KAC00,FAR03} where 
an overly optimistic limit ($ <3 \cdot 10^{-5}$) 
from double beta decay experiments was used.     

For "non-ordinary'' Majoron models, our new limits on $\langle g_{ee} \rangle$ 
are a few times better than reported in \cite{Gunther96,Arn678,Dan03}.
Analysis of the results documented above shows that the best 
limits on the coupling constant for decays with Majoron
emission $(n=3)$ were obtained in the measurement 
of $^{100}\rm Mo$ and for $n=7$ in the measurement of 
$^{82}\rm Se$. For the decay with $n=2$ limit on string scale M can be 
established at the level of $M>1$ TeV (see \cite{Moh00}).

\section{Conclusion}
Improved limits on different Majoron decay modes of $^{100}\rm Mo$ and 
$^{82}\rm Se$ have been obtained. The most stringent limits on 
the Majoron to neutrino
coupling constants have been established. 
Data collection is continuing
and the sensitivity of the NEMO 3 experiment will be increased in the next five years. 
In particular, we hope to improve our knowledge
of detector response function and clarify the situation with the low energy
portion of $^{100}\rm Mo$ spectrum. 
Of course, a much better sensitivity 
($\sim 10^{-5}$  for "ordinary" Majoron) 
will be reached in the next generation double beta decay experiments
(see, for example, review \cite{BAR04a}).

\section*{Acknowledgement}
The authors would like to thank the Modane Underground Laboratory 
staff for their technical assistance in running the experiment. 
Portions of this work were supported by a grant from INTAS (no 03051-3431), 
a NATO grant (PST CLG 980022) and grant from Grant Agency of the Czech Republic 
(202/05/P239).

 --------------------------------------------------------------

\newpage

\newpage

\newpage

\newpage

\newpage

\begin{table}[ht]
\label{Table1}
\caption{Different Majoron models according to \cite{Bamert95,Hirsch96}.
The mode IIF and ''bulk`` correspond to the model \cite{Carone93} and 
\cite{Moh00} respectively.}
\vspace{1cm}
\begin{center}
\begin{tabular}{cccccc}
\hline
Case & Decay mode & Goldstone boson & L & n & Matrix element\\
\hline
IB & $2\beta\chi^{0}$ & no & 0 & 1 & $M_F-M_{GT}$ \\
IC & $2\beta\chi^{0}$ & yes & 0 & 1 & $M_F-M_{GT}$ \\
ID & $2\beta\chi^{0}\chi^{0}$ & no & 0 & 3 & $M_{F\omega^2}-M_{GT\omega^2}$ \\
IE & $2\beta\chi^{0}\chi^{0}$ & yes & 0 & 3 & $M_{F\omega^2}-M_{GT\omega^2}$ \\
IIB & $2\beta\chi^{0}$ & no & -2 & 1 & $M_F-M_{GT}$ \\
IIC & $2\beta\chi^{0}$ & yes & -2 & 3 & $M_{CR}$ \\
IID & $2\beta\chi^{0}\chi^{0}$ & no & -1 & 3 & $M_{F\omega^2}-M_{GT\omega^2}$ \\
IIE & $2\beta\chi^{0}\chi^{0}$ & yes & -1 & 7 & $M_{F\omega^2}-M_{GT\omega^2}$ \\
IIF & $2\beta\chi^{0}$ & gauge boson & -2 & 3 & $M_{CR}$ \\
\hline
''bulk`` & $2\beta\chi^{0}$  & bulk field  &0  & 2 & -- \\
\hline
\end{tabular}
\end{center}
\end{table}

\clearpage

\begin{table}[ht]
\label{Table2}
\caption{Summary of the best results on the $2\beta\chi^{0}$ decay
with $n=1$. All
limits are presented at the 90\% CL. The dispersion of $\langle g_{ee} \rangle$ values is
due to uncertainties in the NME calculation. The NME from the following 
works were used, $3^{rd}$ column: 
$^{48}$Ca - \cite{Retamosa95,Caurier99},
$^{150}$Nd - \cite{SIM99,Hirsch95},
and others - \cite{SIM99,STO01,CIV03}; $4^{th}$ column: \cite{ROD05}.}
\vspace{1cm}
\begin{center}
\begin{tabular}{cccc}
\hline
Nucleus & $T_{1/2}$, y &  \multicolumn{2}{c}{$\langle g_{ee} \rangle\cdot10^{4}$} \\
\hline
$~^{48}$Ca & $>7.2\cdot10^{20}$ \cite{Barabash89} & $<12$\\
$~^{76}$Ge & $>6.4\cdot10^{22}$ \cite{KLA01} & $<(1.2-3.0)$ & $<(1.9-2.3)$\\
$~^{82}$Se & $>1.5\cdot10^{22}$  & $<(0.66-1.4)$ & $<(1.2-1.9)$\\
& (this work) \\
$~^{96}$Zr & $>3.5\cdot10^{20}$ \cite{Arnold99} & $<(3.6-10)$ & $<(35-378)$\\
$~^{100}$Mo & $>2.7\cdot10^{22}$  & $<(0.4-0.7)$ & $<(1.7-1.8)$\\
&(this work)\\
$~^{116}$Cd & $>8\cdot10^{21}$ \cite{Dan03} & $<(1.0-2.0)$ & $<(2.8-3.3)$\\
$~^{128}$Te & $>2\cdot10^{24}$ (geochemical)\cite{MAN91} & $<(0.7-1.6)$ & $<(1.9-2.4)$\\
$~^{130}$Te & $>3.1\cdot10^{21}$\cite{ARNA02} & $<(1.5-4.1)$ & $<(4.7-5.7)$\\
$~^{136}$Xe & $>7.2\cdot10^{21}$ \cite{Luescher98} & $<(1.0-7.4)$ & $<(5.1-6.6)$\\
$~^{150}$Ne & $>2.8\cdot10^{20}$ \cite{Silva97} & $<(2.5-5.5)$ & $<(3.8-4.8)$\\
\hline
\end{tabular}
\end{center}
\end{table}

\clearpage

\begin{table}[ht]
\label{Table_Prob1}
\caption{ Limits on $T_{1/2}$ at 90\% CL for decays with Majoron emission, 
estimated with
likelihood function.}
\vspace{0.5cm}
\begin{center}
\begin{tabular}{ccc|cc}
\hline
Nucleus & $~^{100}$Mo & $~^{82}$Se& \multicolumn{2}{|c}{ Best limits from previuos experiments }  \\
  &         &      &   $^{100}\rm Mo$ & $^{82}\rm Se$ \\
\hline
$n=1$ & $>2.7\cdot10^{22}$ & $>1.5\cdot10^{22}$ & $>5.8 \cdot 10^{21}$ \cite{FUS02}
  &$>2.4 \cdot 10^{21}$ \cite{Arn678}  \\
$n=2$ & $>1.7\cdot10^{22}$ & $>6.0\cdot10^{21}$ & $>3.0\cdot10^{20}$ \cite{BAR04a}  &
$>1\cdot10^{21}$ \cite{BAR04a}\\
$n=3$ & $>1.0\cdot10^{22}$ & $>3.1\cdot10^{21}$ & $1.6 \cdot 10^{20}$ \cite{Arn678} & 
$6.3 \cdot 10^{20}$ \cite{Arn678} \\
$n=7$ & $>7\cdot10^{19}$ & $>5.0\cdot10^{20}$ & $4.1 \cdot 10^{19}$ \cite{Arn678} & 
$1.1 \cdot 10^{19}$ \cite{Arn678}\\
\hline
\end{tabular}
\end{center}
\end{table}

\clearpage

\begin{table}[ht]
\label{Table_Matr}
\caption{The QRPA nuclear matrix elements for
$~^{100}$Mo and $~^{82}$Se.}
\vspace{0.5cm}
\begin{center}
\begin{tabular}{cccc}
\hline
Nucleus & $M_F-M_{GT}$ & $M_{CR}$ & $M_{F\omega^2}-M_{GT\omega^2}$ \\
\hline
$~^{82}$Se & 2.63-5.60 \cite{SIM99,STO01,CIV03} & 0.14-0.44 \cite{Gunther96,BAR97} 
& $10^{-3}$ \cite{Gunther96} \\
$~^{100}$Mo  & 2.97-5.37 \cite{SIM99,STO01,CIV03} & 0.16-0.44 \cite{Gunther96,BAR97} 
& $10^{-3}$ \cite{Gunther96} \\
\hline
\end{tabular}
\end{center}
\end{table}

\clearpage

\begin{table}[ht]
\label{Table_Phase}
\caption{ Phase-space integrals ($G$ [y$~^{-1}$]) for different nuclei and
models of decay (from \cite{SUH98} for n = 1 and from \cite{Gunther96} for n = 3,7.)}
\vspace{0.5cm}
\begin{center}
\begin{tabular}{ccccc}
\hline
Nucleus & $2\beta\chi^{0}$, $n=1$ & $2\beta\chi^{0}$, $n=3$ &
$2\beta\chi^{0}\chi^{0}$,$n=3$ & $2\beta\chi^{0}\chi^{0}$, $n=7$  \\
\hline
$~^{82}$Se
& $4.84\cdot10^{-16}$ & $3.49\cdot10^{-18}$ & $1.01\cdot10^{-17}$ & 
$7.73\cdot10^{-17}$\\
$~^{100}$Mo
& $8.23\cdot10^{-16}$ & $7.28\cdot10^{-18}$ & $1.85\cdot10^{-17}$ & 
$1.54\cdot10^{-16}$\\
\hline
\end{tabular}
\end{center}
\end{table}

\clearpage

\begin{table}[ht]
\label{Table_Const}
\caption{ Limits on the Majoron coupling constant $\langle g_{ee} \rangle$ at the $90\%$ CL
for $~^{100}$Mo and $~^{82}$Se.}
\vspace{0.5cm}
\begin{center}
\begin{tabular}{ccccc}
\hline
model & mode & n & $~^{82}$Se & $~^{100}$Mo \\
\hline
IB & $2\beta\chi^{0}$ & 1
& $(0.66-1.7)\cdot10^{-4}$  & $(0.4-1.8)\cdot10^{-4}$\\
IC & $2\beta\chi^{0}$ & 1
& $(0.66-1.7)\cdot10^{-4}$& $(0.4-1.8)\cdot10^{-4}$\\
IIB & $2\beta\chi^{0}$ & 1
& $(0.66-1.7)\cdot10^{-4}$& $(0.4-1.8)\cdot10^{-4}$\\
\hline
ID & $2\beta\chi^{0}\chi^{0}$ & 3
& 2.4& 1.5\\
IE & $2\beta\chi^{0}\chi^{0}$ & 3
& 2.4& 1.5\\
IIC & $2\beta\chi^{0}$ & 3
& 0.022-0.068& 0.0088-0.024\\
IID & $2\beta\chi^{0}\chi^{0}$ & 3
& 2.4& 1.5\\
IIF & $2\beta\chi^{0}$ & 3
& 0.022-0.068& 0.0088-0.024\\
\hline
IIE & $2\beta\chi^{0}\chi^{0}$ & 7
& 1.3& 3.2\\

\hline
\end{tabular}
\end{center}
\end{table}

\end{document}